\documentclass[
    ,final            
  ]
  {aipproc}

\layoutstyle{6x9}

\begin{document}

\title{Consistent relativistic Quantum Theory for systems/particles 
described by non-Hermitian Hamiltonians and Lagrangians}

\author{Frieder Kleefeld}{
  address={Centro de F\'{\i}sica das Interac\c{c}\~{o}es Fundamentais (CFIF), Instituto Superior T\'{e}cnico,\\
Edif\'{\i}cio Ci\^{e}ncia, Piso 3, Av. Rovisco Pais, P-1049-001 LISBOA, Portugal\\
e-mail: kleefeld@cfif.ist.utl.pt \\
\ \\
$\ldots$ to Manuela, Alexander and Sara $\ldots$}
}

\begin{abstract}
A causal, non-Hermitian, renormalizable, local, unitary and Lorentz convariant formulation of Quantum Theory (QT) (= Quantum Mechanics (QM) and Quantum Field Theory (QFT)) is developed which is free of formalistic problems we face in the commonly used Hermitian formulation of QT. Side effects of the new formulation of QT are the derivation of a consistent (anti)causal neutrino Lagrangian, the enrichment of chiral symmetries, the removal of the Dirac sea, the separation of positive and negative energy states including a reformulation of the anti-particle concept and a critical analysis of the concept of probability currents. In a first step we apply the new formulation of QT to establish
a relation between perturbative Quantum Chromodynamics (QCD) and the (axial)vector meson extended Quark-Level Linear Sigma Model (QLLSM) at high energies.  
\end{abstract}

\maketitle


\section{Introduction to (anti)causal Quantum Theory}

It is well known --- that the present Hermitian formulation of QT (see e.g.\ \cite{araki1}) gives rise to
a variety of formalistic --- unresolved --- problems and paradoxa. Without going into details\footnote{Some examples: problems in the probability concept of QT \cite{mueckenheim1}, the EPR paradoxon \cite{afriat1}, the Klein paradoxon, Zitterbewegung, supercritical fields and the problem of gauge fixing \cite{greiner1}, the problem of bifurcation or compositeness in QM (noted e.g. by T.D.\ Lee \cite{lee1}), in QFT (noted e.g.\ by L.D.\ Landau (p.\ 18 in \cite{fradkin1} and references therein)), the problem of regularization and anomalies (e.g. \cite{zinn1}).} we want to stress that such conceptual shortcomings get more pronounced, if we try to apply the Hermitian QT (HQT) to ``non-Hermitian'' (e.g.\ absorptive, resonant, bifurcating, CP-violating, $\ldots$) problems. Simple consequences are e.g.\ the loss of locality, causality, stability of the effective action and the positivity of spectral functions. We will show that many of these problems can be understood and circumvented by treating QT in strict sense (anti)causal. In standard HQT --- as we teach it to the students --- causality is implemented by analytical continuation of Green's functions or propagators according to (anti)causal boundary conditions to be imposed. What we don't teach the students is that this analytical continuation (usually performed in generating functionals) not only spoils the unitarity of the HQT (due to an infinitesimally non-Hermitian effective action, i.e. a non-unitary S-matrix), yet also has severe negative implications on its consistency when afterwards demanding field operators or wavefunctions to stay Hermitian yielding a lot of paradoxa mentioned above\footnote{There are also non-Hermiticities inserted in QT by hand, which may lead to serious problems, if one treats the field operators or wavefunctions furtheron Hermitian. Examples are complex mixing angles \cite{benayoun1}, resonance propagators \cite{kniehl1}, CP-violating phases \cite{espriu1,gouvea1} in high energy, hadronic or solid-state physics, imaginary chemical potentials e.g.\ in QCD \cite{delia1}, optical potentials in nuclear physics, $\ldots$}.
E.g.\ in Refs.\ \cite{kniehl1,espriu1} it is noted that that decay rates of the electroweak gauge Bosons cannot be described in a gauge-invariant manner, if they are not determined on the complex mass shell of the respective gauge Bosons. 

\emph{What is to be understood by a complex mass shell?} Let us consider for simplicity a neutral non-interacting Klein-Gordon (KG) field $\phi (x)$. In QFT the real mass $m$ of $\phi (x)$ has to be analytically continued according to $m \rightarrow m -  i\,\varepsilon$, yielding a \emph{complex(valued)} mass $M = m - \frac{i}{2} \Gamma$ with $\Gamma \rightarrow \varepsilon = + 0$. For resonances like the electroweak gauge Bosons the width $\Gamma$ will be even finite. As the complex mass $M$ is entering the analytically continued \emph{causal} generating functional the respective KG field will obey the classical equation of motion $((i\,\partial)^2 - M^2) \,\phi (x) = 0$. A neutral spin 1/2 Fermion will correspondingly fulfil the respective causal Dirac equation $( i \!\not\!\partial - M) \psi (x) = 0$. As the imaginary part of $M$ is finite the \emph{causal} KG and Dirac equations have to be solved by a \emph{Laplace-transform} and not by a simple \emph{Fourier-transform} as taught in textbooks. Consequently e.g.\ the neutral KG field $\phi(x)$ will be \emph{complex(valued)} and \emph{not} real or Hermitian as stated in textbooks\footnote{The mass shell condition in 4-momentum space $p^2 = M^2$ will therefore be \emph{complex} yielding the singular points $p^0 = \pm \,\omega(\vec{p})$ (with $\omega(\vec{p}):= \sqrt{\vec{p}^{\,2} + M^2}$ and $\omega(\vec{0}):= M$) in the complex $p^0$-plane. Hence in performing the Laplace-transform the solution of the causal KG (or Dirac) equation can be decomposed by $\phi(x) = \phi^{(+)}(x)+\phi^{(-)}(x)$ (or $\psi(x) = \psi^{(+)}(x)+\psi^{(-)}(x)$), while $\phi^{(\pm)}(x)$ (or $\psi^{(\pm)}(x)$) is associated with $p^0 =\pm \,\omega(\vec{p})$, respectively.}${}^{,}$\footnote{Hermitian conjugation of the causal KG equation will yield what we call \emph{anticausal} KG equation, i.e.\  $((i\,\partial)^2 - M^{+ \, 2}) (\phi(x))^+ = 0$. Due to its properties we call the \emph{anticausal} KG field a KG \emph{hole field} \cite{kleefeld1}. As $M^+ \not= M$, it is clear that $(\phi(x))^+ \not= \phi (x)$ and $\phi^{(+)}(x)\not=(\phi^{(-)}(x))^+$. In QM \emph{causal} and \emph{anticausal} states are called \emph{Gamow states} and \emph{anti-Gamow states}, respectively (see e.g.\ Refs.\ \cite{kaldass1,moiseyev1}).}. More explicitly we obtain\footnote{We introduce a Dirac spinor $u(p,s) \equiv v(-p,s)$) fulfilling the equation $(\not\!\!p - \sqrt{p^2}\;)\, u(p,s) = 0\,$ or, equivalently,  the \emph{transposed} equation $\overline{u^c}(p,s) \; ( - \not\!\!p -  \sqrt{p^2} \; ) \; = \; 0$. Important identities: $u^c(p,s) = u(-\,p^\ast ,s) = v(p^\ast ,s)$ and 
sgn[Re$(p^0)] \; \sum_s u(p,s) \, \overline{v^c}(p, s) =  \; \not\!\!p +  \sqrt{p^2}\; $ (for Re$[p^0]\not=0$). 
Furthermore we may define: \\[1.2mm]
$a(\vec{p}) := a(p) |_{p^0 = \omega(p)}$, $c^+(\vec{p}) := a(-p) |_{p^0 = \omega(p)}$, $b(\vec{p},s) := b(p,s) |_{p^0 = \omega(p)}$, $d^+(\vec{p},s) := b(-p,s) |_{p^0 = \omega(p)}$.}:
\begin{eqnarray} \phi (x) & = & \int \! \! \frac{d^4 p}{(2\pi )^3} \; \delta(p^2 - M^2)\; a \, (p ) \, e^{ - \, i p x} \stackrel{!}{=} \sum\limits_{\pm} \int \! \! \frac{d^3 p}{(2\pi )^3 \, 2 \omega (\vec{p})}
\,  a \, (\pm p ) \, e^{ \mp \, i p x} \Big|_{p^0 = \omega(p)} \nonumber \\
\psi (x) & = & \sum\limits_s \int \! \! \frac{d^3 p}{(2\pi )^3 \, 2 \omega (\vec{p})}
\,
\Big[ \, 
 b \, (p,s) \, u (p,s) \, e^{ - \, i p x} + 
 b \, (-p,s) \, u (- p,s) \; e^{i p x}
\,\Big] \Big|_{p^0 = \omega(p)} \nonumber 
\end{eqnarray}
The symbolic $\delta$-distribution ``$\delta(p^2 - M^2)$'' should be understood in the sense of residuum calculus and Cauchy's theorem and finds its solid mathematical basis in a formalism worked out by N.\ Nakanishi \cite{nakanishi1,nakanishi2}. The definition of spinors (and polarization vectors) requires the introduction of a Lorentz boost for objects with \emph{complex} mass $M$~\footnote{A Lorentz transformation (LT) $\Lambda^{\mu}_{\,\;\nu}$  is defined by $\Lambda^{\mu}_{\,\;\rho} \; g_{\mu\nu} \; \Lambda^{\nu}_{\,\;\sigma} = g_{\rho\sigma}$ for any metric $g_{\mu\nu}$. Let $n^{\mu}$ be a timelike unit 4-vector $(n^2 = 1)$ and $\xi^{\mu}$ an \emph{arbitrary complex} 4-vector with $\xi^2 \not= 0$. The 4 independent LTs relating $\xi^{\mu}$ with its ``restframe'' (i.e. $\xi^{\, \mu} = \Lambda^{\,\mu}_{\,\;\,\nu} (\xi) \; n^{\,\nu} \sqrt{\xi^2}$ and $n_{\,\nu} \sqrt{\xi^2} = \xi_{\, \mu} \, \Lambda^{\,\mu}_{\,\;\,\,\nu} (\xi)$) are:
\[ \Lambda^{\,\mu}_{\,\;\,\nu} (\xi) =  \pm \, \left\{ g^{\,\mu}_{\,\;\,\rho} - \, \frac{\sqrt{\xi^2}}{\sqrt{\xi^2} \mp \xi \cdot n} \, \Big[ n^{\,\mu} \mp \frac{\xi^{\,\mu}}{\sqrt{\xi^2}} \Big] \Big[ n_{\,\rho} \mp \frac{\xi_{\,\rho}}{\sqrt{\xi^2}} \Big] \, 
\right\} \,  P^{\,\rho}_{\,\;\,\nu}  \quad \makebox{and} \quad \Lambda^{\,\mu}_{\,\;\,\rho} (\xi)\, P^{\,\rho}_{\,\;\,\nu} \]
(with $P^{\,\mu}_{\,\;\,\nu} := 2\, n^{\,\mu} \, n_{\,\nu} - \, g^{\,\mu}_{\,\;\,\nu}\;$ = reflection matrix). Lorentz covariance of the Dirac equation requires $S^{ -1} (\Lambda (p)) \, \gamma^{\,\mu} \, S (\Lambda (p)) \, = \, \Lambda^{\,\mu}_{\,\,\;\nu} (p) \; \gamma^{\,\nu}$ for $u(p,s) = S(\Lambda(p)) \; u(\sqrt{p^2} \; n,s)$. The metric $(+,-,-,-)$ yields
\[ u(p,s) \big|_{\mbox{Re$[p^0]>0$}} = \frac{\not\!\!p +  \sqrt{p^2}}{\sqrt{2\,\sqrt{p^2} \; (p^0 + \sqrt{p^2})}} \; \; u(\sqrt{p^2},\vec{0},s) \stackrel{p^0 = \omega(\vec{p})}{\longrightarrow} \frac{\not\!\!p + M}{\sqrt{2\,M \; (\omega(\vec{p}) + M)}} \; \; u(M,\vec{0},s) \; .\]
Similarly construct $v(p,s)$, $u^c (p,s)$, $v^c (p,s)$, $\bar{u} (p,s)$, $\bar{v} (p,s)$, $\overline{u^c}(p,s)$ and $\overline{v^c}(p,s)$ for Re$[p^0]>0$!}. 
Hermiticity is also broken in the causal time-dependent Schr\"odinger equation:
\begin{equation}  i \,\partial_t \, \psi (x) = \Big(- \frac{1}{2\, M} \, \stackrel{\rightarrow}{\nabla}{}^2 + V_C(x) \Big) \, \psi (x) = H_C \; \psi (x) \, .
\end{equation}
Due to the fact that $M$ is complex and the \emph{causal potential} $V_C(x) (\not= V^+_C(x) =: V_A(x)$) is essentially the Laplace-transform of a causal propagator we notice that the \emph{causal Hamilton operator $H_C$} entering the causal Schr\"odinger equation is not Hermitian, i.e. $H_C \not = H^+_C = : H_A$ (while $H_A$ is called the \emph{anticausal Hamilton operator})\footnote{For $M = m - i\,\varepsilon$ we may call $H_C$ \emph{quasi-Hermitian}, if $V_C(x)$ is in a similar sense quasi-Hermitian.}. As $H_C$ is not Hermitian its right ($\psi(x)$) and left eigenfunctions ($\tilde{\psi}(x)$) are \emph{not} related by Hermitian conjugation\footnote{Causal ``bra's'' (``out-states'') are \emph{not} Hermitian conjugate to causal ``ket's'' (``in-states'') ($\left<\tilde{\psi}\right| \not= \left|\psi\right>^+$)!}. $\tilde{\psi}(x)$ may be called \emph{adjoint causal} Schr\"odinger wavefunction fulfilling the \emph{adjoint causal Schr\"odinger equation} $ -\, i \,\partial_t \, \tilde{\psi} (x) = \tilde{\psi}(x) \,H_C$.

Pars pro toto we are now going to discuss further aspects in the context of the (anti)causal KG field. For convenience we define the quantity $\sqrt{Z} \,:= |M|/M$ and the field $\varphi(x) := \phi (x)/\sqrt{N\, Z}$. The non-Hermitian \emph{causal (anticausal)} Lagrangian ${\cal L}_{C} (x)$ (${\cal L}_{A} (x)$) with ${\cal L}_{C} (x) = ({\cal L}_{A} (x))^+$ yielding the \emph{causal (anticausal)} KG equation is:
\begin{equation} {\cal L}_{C} (x) = \frac{1}{2\,N} \, \Big( (\partial\,\phi (x) )^2  - \, M^2 \, \phi (x)^2 \, \Big) = \frac{1}{2} \, \Big( Z \, (\partial\,\varphi (x) )^2  - \, |M|^2 \, \varphi (x)^2 \, \Big) \, . \end{equation}
The complex constant $N$ is just the square of an overall normalization of the causal field $\phi (x)\;$\footnote{We may call $\sqrt{N}$ the \emph{causal norm} of a state/field.}. Typically we choose $N$ to be  real and positive (e.g. $N=1$) for (anti)causal fields, while we call fields for which $N$ is of opposite sign (e.g. $N= -1$) (anti)causal \emph{ghost fields}. 
We see that --- in the context of a HQT --- the quantity $\sqrt{Z}$ gets --- with respect to $\varphi (x)$ --- the meaning of a traditional \emph{wavefunction renormalization}. Due to its definition we know that $|Z|=1$. For $Z = \exp (i\, \varepsilon )$ (with $\varepsilon = + 0$) the causal fields $\phi(x)$ and $\varphi(x)$ may be called \emph{asymptotic stable}, as they can represent asymptotic stable systems\footnote{Such fields appear in the context of the $S$-matrix as in- and out-states.}. If Im$[\sqrt{Z}\,]>0$ and non-infinitesimal, the width $\Gamma>0$ will also be finite. Then the fields $\phi(x)$ and $\varphi(x)$ may be called \emph{non-asymptotic} or \emph{unstable}, as they represent \emph{intermediate states} with a finite life time\footnote{This is rather consistent with the need of \emph{complex wavefunction renormalizations} stressed in \cite{kniehl1,espriu1}.}. In the context of HQT it is the general belief (see e.g.\ Ref.\ \cite{dass2}) that composite states are considered to be states without kinetic energy, i.e.\ with vanishing wavefunction renormalization, i.e.\ $Z\rightarrow 0$. In a (anti)causal QT \emph{this is not possible} due to the constraint $|Z|=1$. Nevertheless we may declare states/fields to be \emph{composite}, when Re$[\sqrt{Z}\,]\rightarrow 0$. A transition from Im$[\sqrt{Z}\,]>0$ to Im$[\sqrt{Z}\,]<0$ will bring us from causal to anticausal fields. The Hermitian situation Im$[\sqrt{Z}\,]=0$ describes Hermitian \emph{acausal} fields, whose propagators are half causal and half anticausal ($=$ principal value prescription). For historical reasons we will call such (``standing wave'', i.e. non-propagating) states  \emph{shadow states} \cite{nakanishi3,stapp1}\footnote{These shadow states shouldn't be confused with the shadow states addressed e.g. by T.\ Regge \cite{regge1}!}. Finally --- in accordance with earlier discussions in the literature on HQT --- we want to call here Re$[\sqrt{Z}\,]$ the \emph{traditional norm} of a state/field. With the help of the preceding discussion we are now in the position to consider selected historical difficulties in the formalistic developement of a Hermitian \emph{and} causal QT which persist upto now. 

In a first step we want to address here the ``bifurcation'' or ``compositeness problem''. In 1954 T.D.\ Lee \cite{lee1} (see also \cite{heisenberg1,nakanishi1,nakanishi2,hittner1,casagrande1,rabuffo1}) observed in the \emph{Hermitian} ``Lee-model''  that the renormalization of the model required a purely imaginary unrenormalized coupling constant to obtain a finite real renormalized coupling constant, and that beyond a certain critical value of the coupling constant, at which the wavefunction renormalization went to zero (i.e. $Z\rightarrow 0$), the \emph{Hermitian} Hamiltonian developed \emph{Hermitian conjugate complex pairs of eigenvalues} for pairs of \emph{zero traditional norm eigenstates} to be considered as \emph{metric partners} \footnote{A similar singular phenomenon is known in perturbative QED as ``Landau pole'' \cite{tarrach1}. As a non-perturbative phenomen bifurcation has been observed e.g.\ in QED in Dyson-Schwinger calculations at a critical value of $\alpha_{cr}>0$ \cite{ahlig1,sauli1,maris1} (signalled by numerical instabilities and hugely enhanced calculation time or complex branchcuts), in variational methods at $\alpha_{cr} \le 0$ \cite{schreiber1,markushin1} and on the lattice \cite{helen1}.} forming a \emph{biorthogonal basis} \cite{mostafazadeh1} \footnote{T.D.\ Lee surprised that a purely imaginary (unrenormalized) coupling constant enters a HQT remarked: ``This difficulty may, however, be overcome by a modification of the present rules of quantum mechanics''.}. The problem of the treatment of zero traditional norm states and negative causal norm states (``ghosts'') had --- since Dirac \cite{dirac1,dirac2} --- induced a lot of theoretical activities which culminated in those times among other things in the concept of an indefinite metric and a metric operator (for details see e.g.\ Refs. \cite{nakanishi2,rabuffo1,pauli1,mostafazadeh1}). The application of such concepts has faced upto now enormous difficulties in the context of HQT, mainly because most calculations involved a (blind) \emph{mixture} of purely Hermitian states (``shadow states'') and zero traditional norm states (``(anti)causal states'') and their interactions. As we will argue in the following \emph{a consistent treatment within a (anti)causal QT requires the sole existence of causal and anticausal states (consisting of a special superposition of ``shadow states'') such that both classes states --- causal and anticausal --- don't interact !} As a consequence the shadow states interact in a very restricted way which is governed by the Lorentz covariance of the theory \cite{nakanishi3,stapp1}. Furthermore, as $|Z|=1$, intermediate unstable fields cannot be easily integrated out from a (anti)causal Lagrangian. We also have to realize that the restriction to a ``Hermitian'' Lehmann-K\"all\'{e}n spectral representation \cite{lehmann1,schwinger1} (in which spectral functions are real functions of real invarinant mass variables) is an inadmittable corset imposed by HQT, which does not allow QT to bifurcate\footnote{This explains partially the technical and numerical difficulties we have in describing QED close to the Landau pole or QCD close to the chiral or (de)confinement phase transition(s).}. In (anti)causal QT spectral functions are complex functions of complex invariant mass variables even for asymptotic states \cite{spitzenberg1,markushin1}.  

Secondly we want to show an obvious inconsistency in the probability concept of traditional QT. The way students learn to derive probability currents in QT is to subtract classical equations of motion and its \emph{Hermitian conjugate} yielding in (anti)causal QT the following continuity-like equations for neutral KG, Dirac and Schr\"odinger theory\footnote{For simplicity we display here only the quasi-real case with $M = m - i\varepsilon$ and $\varepsilon = + 0$.}:
\begin{eqnarray} \lefteqn{\partial_\mu [ \, \phi^+ (x) \; \partial^\mu \phi (x)  -  (\,\partial^\mu \phi^+ (x) ) \, \phi(x) \, ]  = 2\, i \, \varepsilon \,\phi^+ (x) \, \phi (x) } \\[2mm]
\lefteqn{\partial_\mu [ \, i\, \overline{\psi} (x) \; \gamma^\mu\;  \psi (x) \, ]  =   -\, 2\, i \, \varepsilon \,\bar{\psi} (x) \, \psi (x) } \\
\lefteqn{i \partial_t \, [ \, \psi^+ (x) \, \psi (x) \, ] + \frac{1}{2\, m} \!\stackrel{\rightarrow}{\nabla} \!\cdot \,[ \, \psi^+ (x) \stackrel{\rightarrow}{\nabla} \psi (x) - ( \,\stackrel{\rightarrow}{\nabla} \psi^+ (x) \, ) \, \psi (x)\, ] =} \nonumber \\
& = &
\psi^+ (x) \, [ \, V_C(x) - V_C^+(x) \,] \, \psi (x) - \frac{i\,\varepsilon}{2\, m} \,[ \, \psi^+ (x) \stackrel{\rightarrow}{\nabla}  {}^2 \, \psi (x) + ( \, \stackrel{\rightarrow}{\nabla}  {}^2 \, \psi^+ (x) \, ) \, \psi (x) \, ]
\end{eqnarray}
We see that \emph{none} of the three currents will be conserved, if we use a complex mass $M$ and a causal potential $V_C(x)$. This means for the Schr\"odinger theory that $|\psi(x)|^2$ \emph{cannot} be considered to be the density of a conserved current (being a minimal requirement in the context of probability theory) and hence should \emph{not} be interpreted as a probability density as originally done by M.\ Born \cite{pais1}! Alternaltively, if we perform the subtraction of the causal KG equation for $\phi^{(\pm)}$ (x) (Dirac equation for $\psi^{(\pm)}(x)$, Schr\"odinger equation for $\psi(x)$, respectively) and the \emph{transposed} (or adjoint) equation for $\phi^{(\mp)}(x)$ ($\overline{\psi^{(\mp)c}}(x)$ \footnote{We adopt the standard notation $\overline{\psi^c} (x) := \psi^{\,T} (x) \; C$ (with $C = i \,\gamma^{\,2} \, \gamma^{\,0}\,$) and $\overline{\psi}(x) := \psi^+(x) \; \gamma^{\,0}$.} and $\tilde{\psi}(x)$, respectively) we will obtain the following continuity equations:
\begin{eqnarray} \partial_\mu [ \, \phi^{(\mp)} (x) \; \partial^\mu \phi^{(\pm)} (x)  - (\partial^\mu \phi^{(\mp)} (x)) \, \phi^{(\pm)} (x) \, ]  & = & 0  \label{boscur1} \\[2mm]
\partial_\mu [ \, i\, \overline{\psi^{(\mp)\, c}} (x) \; \gamma^\mu\;  \psi^{(\pm)} (x) \, ]  & = & 0 \label{fercur1} \\
 i \partial_t \, [ \, \tilde{\psi} (x) \, \psi (x) \, ] + \frac{1}{2\, M} \!\stackrel{\rightarrow}{\nabla} \!\cdot \,[ \, \tilde{\psi} (x) \stackrel{\rightarrow}{\nabla} \psi (x) - (\, \stackrel{\rightarrow}{\nabla} \tilde{\psi} (x) \, ) \, \psi (x) \, ] & = & 0 
\end{eqnarray}
The respective \emph{non-vanishing}\footnote{Even in the case of the neutral KG field this current is not vanishing!} causal currents are \emph{conserved} even for \emph{arbitrary} complex mass $M$ and causal potential $V_C(x)$\footnote{The respective isospin currents for neutral KG and Dirac fields vanish (due to a tricky cancellation of the underlying currents of Eqs. (\ref{boscur1}) and (\ref{fercur1})), i.e.\ $\phi (x) \; \partial^\mu \phi (x)  -  (\partial^\mu \phi (x)) \,\phi(x) = 0$ and $i\, \overline{\psi^{\, c}} (x) \; \gamma^\mu\;  \psi (x)= 0$, if we treat $\phi (x)$ as a \emph{commuting} field and $\psi (x)$ as a \emph{anticommuting Grassmann} field!}. Within the (anti)causal framework charge, standard gauge invariance and the anti-particles are introduced according to the isospin concept, i.e. by considering $N$ causal fields (i.e. $\phi_r(x)$ or $\psi_r(x)$ with $r=1,\ldots,N$) of equal complex mass $M$. Unitarity of the theory is restored by considering the Hermitian Lagrangian ${\cal L} (x) = {\cal L}_C (x) + {\cal L}_A (x)$ (or Hamiltonian) containing causal \emph{and} anticausal fields\footnote{I.e.\ a (anti)causal prescription of QT yields a ``doubling'' of degrees of freedom with respect to the Hermitian (acausal) description. Obviously this ``doubling'' commonly performed at finite temperature $T$ for consistency reasons \cite{weldon1} is for causality and unitarity reasons already present at $T=0$.}. For the (anti)causal free KG \cite{nakanishi3,nakanishi2,kleefeld1} and Dirac \cite{kleefeld2,kleefeld3,kleefeld4} theory and e.g.\ for the simple (anti)causal Harmonic oscillator in QM (e.g. \cite{moiseyev1}) we have ($\bar{M} := \gamma_0 \; M^+ \gamma_0$):
\begin{eqnarray} {\cal L} (x) & = & \sum\limits_{r} \;
\frac{1}{2} \Big( (\partial \,\phi_r (x) )^2  - M^2 
(\phi_r (x))^2 +  (\partial \,\phi^+_r (x) )^2  - M^{\ast \,2} 
( \phi^+_r (x))^2 \Big) \label{kglag1} \\ 
{\cal L} (x) & = & \sum\limits_r \frac{1}{2} \, \Big( \overline{\psi_r^c} (x) \, ( \frac{1}{2} \, i \! \stackrel{\;\,\leftrightarrow}{\not\!\partial} \! -  M ) \, \psi_r (x) \; + \; \overline{\psi}_r (x) \, ( \frac{1}{2} \; i \! \stackrel{\;\,\leftrightarrow}{\not\!\partial} \! -  \bar{M} ) \, \psi_r^c (x) \; \Big) \label{diraclag1} \\
\left< z,z^\ast| H |z^\prime,z^{\prime\ast}\right> & = & \Big( - \frac{1}{2 M} \frac{d^2}{dz^2} + \frac{1}{2} \, M \Omega^2 z^2 - \frac{1}{2 M^\ast} \frac{d^2}{dz^{\ast 2}} + \frac{1}{2} \, M^\ast \Omega^{\ast 2} z^{\ast 2}\Big) \! \left< z,z^\ast|z^\prime,z^{\prime\ast}\right> \nonumber
\end{eqnarray}
For $N=1$ Eq.\ (\ref{diraclag1}) is describing the causal Lagrangian of a neutral Fermion (i.e. a Majorana--like neutrino \cite{gouvea1})\footnote{This can be understood from the Grassmann--nature of the Fermion fields as $(\overline{\psi^c} (x) \!\not\!A (x) \, \psi (x))^T = - \, \overline{\psi^c} (x)\! \not\!A (x) \, \psi (x) = 0$ for a ``symmetric'' Abelian/non--Abelian gauge field $A_\mu (x) = (A_\mu (x))^T$. In HQT the kinetic term of such a Majorana--like field would vanish, while in (anti)causal QT it survives!}. For $N=2$ we may define the ``charged'' fields $\psi_\pm (x) := (\psi_1(x)\pm i\psi_2(x))/\sqrt{2}$ (representing e.g.\ a causal positron and electron\footnote{In a isospin--concept (causal) particles and antiparticles (described by $\psi_\pm (x)$) having positive energy and moving forward in time \cite{kleefeld1} possess same parity, while the parity of (anticausal) holes and antiholes (described by $\psi^c_\mp (x)$ bearing negative energy and moving backward in time \cite{kleefeld1} is opposite. This is perfectly possible without spoiling experimentally confirmed results of theoretical calculations in HQFT, e.g.\ meson or positronium spectra, as this type of internal parity is subject to a superselection rule \cite{wick1}.)}) and denote the respective minimally coupled Dirac-like Lagrangian of  such a simply charged Fermion:
\begin{equation} {\cal L} (x) =  \overline{\psi_+^c} (x) ( \frac{1}{2} \, i \!\! \!\stackrel{\;\,\leftrightarrow}{\not\!\partial} \!\!+ g \not\!A (x) -  M ) \, \psi_- (x)  +  \overline{\psi}_- (x) ( \frac{1}{2} \, i \! \!\!\stackrel{\;\,\leftrightarrow}{\not\!\partial} \!\! + g^\ast \gamma^{\,\mu} A^+_{\mu} (x) -  \bar{M} ) \psi_+^c (x)
\end{equation}
being invariant under the local gauge transformations $g \not\!\!A^{\,\prime} = g \not\!\!A + [\not\!\partial , \Lambda (x)]$, $\psi^{\,\prime}_- (x) = \exp(+ i \, \Lambda (x)) \; \psi_- (x)$, $\psi^{\,\prime}_+ (x) = \exp(- i \, (\Lambda (x))^T) \; \psi_+ (x)$ \footnote{Within such an isospin concept we obtain for $A^\mu(x) = 0$ the following continuity equations for $N=2$: 

\makebox[0.5mm]{}$\partial_\mu [ \, i\, \overline{\psi^c_{\mp}} (x) \; \gamma^\mu\;  \psi_{\pm} (x) \, ]  = 0$ (Dirac--field) and  $\partial_\mu [ \, \phi_{\mp} (x) \; \partial^\mu \phi_{\pm} (x)  - (\partial^\mu \phi_{\mp} (x)) \, \phi_{\pm} (x) \, ]  = 0$ (KG field).}. The same holds for a non-Abelian gauge theory, i.e.\ for $A_\mu (x) =  A_\mu^{a} (x)\, \lambda^a/2$ and $\Lambda (x) =  \Lambda^{a} (x)\, \lambda^a/2\;$\footnote{Recently Jun-Chen Su \cite{jun1} showed that a renormalizable Lagrangian for a massive (non-)Abelian vector Boson originally derived by `t Hooft \cite{hoo1} within the context of a Higgs-mechanism persists renormalizable without relying on any Higgs mechanism. It is easy to extend such a  Lagrangian consistently to complex mass vector fields implying that the whole standard renormalization concept of QFT on the basis of gauge invariance is working particularly well for (anti)causal fields of arbitrary complex mass. In this context we have to accept that vector Bosons consist of \emph{three} physical degrees of freedom!}. Causal (anti)particles want to minimize their energy \cite{kleefeld1}, anticausal (anti)holes want to maximize their energy. As long as the vacuum for (Fermionic or Bosonic) (anti) particles and (anti)holes is situated at zero energy and no interactions between (anti)particles and (anti)holes are permitted, positive and negative energy states are nicely separated. There will be no need for any Dirac sea\footnote{Due to its infinite mass a Dirac sea would induce the immediate gravitational collapse of the universe!}, it will be not necessary to forbid the Bose-Einstein condensation of Bosons \cite{dirac1} and there will be no paradoxa like the Zitterbewegung or the Klein paradoxon.  
Finally it is of interest to study the Hermiticity content of (anti)causal fields. Hence we decompose the (anti)causal \emph{neutral} KG fields $\phi^+(x)$, $\phi(x)$ and neutrino fields $\psi^c(x)$, $\psi(x)$ in shadow fields\footnote{This is done by $\phi(x) =: ( \phi_{(1)}(x) + i\, \phi_{(2)} (x))/\sqrt{2}\;$, $\phi^+(x) =: ( \phi_{(1)}(x) - i\, \phi_{(2)} (x))/\sqrt{2}\;$ and $\psi(x) =: ( \psi_{(1)}(x) + i \, \psi_{(2)} (x))/\sqrt{2}$,  $\psi^c(x) =: ( \psi_{(1)}(x) - i \, \psi_{(2)} (x))/\sqrt{2}\,$.} yielding the following Lagrangians:
\begin{eqnarray} {\cal L} (x) & = & 
\frac{1}{2} \Big( (\partial \phi_{(1)} (x) )^2  - \mbox{Re}[M^2]  
(\phi_{(1)} (x))^2 \Big) -  \frac{1}{2} \Big( (\partial \phi_{(2)} (x) )^2  -  \mbox{Re}[M^2]  (\phi_{(2)} (x))^2 \Big) \nonumber  \\ 
 & + & \mbox{Im}[M^2]  \; \phi_{(1)} (x) \,\phi_{(2)} (x) \label{kgshad1} \\[1mm]
{\cal L} (x) & = & \frac{1}{2} \, \overline{\psi}_{(1)} (x) \, \Big( \frac{1}{2} \; i \! \stackrel{\;\,\leftrightarrow}{\not\!\partial} \! -  \,\mbox{Re}[M] \Big) \, \psi_{(1)} (x) - \, \frac{1}{2} \; \overline{\psi}_{(2)} (x) \, \Big( \frac{1}{2} \; i \! \stackrel{\;\,\leftrightarrow}{\not\!\partial} \! - \, \mbox{Re}[M] \Big) \, \psi_{(2)} (x) \nonumber  \\
 & + &   \frac{1}{2} \; \mbox{Im}[M] \; \Big( \overline{\psi}_{(2)} (x) \;\psi_{(1)} (x) +  \overline{\psi}_{(1)} (x) \;  \psi_{(2)} (x) \Big) \label{diracshad1} 
\end{eqnarray} 
We observe that the diagonal (anti)causal Lorentz covariant Lagrangians Eqs.\ (\ref{kglag1}) and (\ref{kgshad1}) are \emph{non-diagonal} in the shadow fields and that the two (anti)causal fields consist of one shadow field with \emph{positive} and one with \emph{negative} norm\footnote{N.\ Nakanishi \cite{nakanishi2,nakanishi3} was the first who investigated the KG Lagrangians Eqs.\ (\ref{kglag1}) and (\ref{kgshad1}) in more detail. Unfortunately he lost --- as we think --- the interest in his ``Complex-Ghost Relativistic Field Theory'' due to singularities he faced by allowing interactions between causal and anti-causal fields!}. Lagrangian Eq.\ (\ref{diracshad1}) is particularly useful to study chiral symmetry in the context of (anti)causal QT. It is straight forward to show that a simultaneous chiral transformation in of the shadow fields according to\footnote{We mention that $\gamma_C := -\, i \, C\;$ behaves in many aspects like $\gamma_5$ --- particularly in (anti)causal QT.} $\psi_{(1)} (x) \rightarrow \exp (i\,\gamma_5\, \alpha) \, \psi_{(1)} (x)$ and $\psi_{(2)} (x) \rightarrow \exp (- i\,\gamma_5\, \alpha) \, \psi_{(2)} (x)$ will yield the continuity-like equation\footnote{Of course there exist also the respective Hermitian conjugate and transposed continuity-like equations!} $\partial_\mu [ \, \overline{\psi} (x) \; \gamma^{\,\mu} \, \gamma_5  \psi (x) \, ] \propto \mbox{Re}[M]$, while the chiral rotation $\psi (x) \rightarrow \exp (i\,\gamma_5\, \alpha) \, \psi (x)$ and $\psi^c (x) \rightarrow \exp (i\,\gamma_5\, \alpha) \, \psi^c (x)$ in Eq.\ (\ref{diraclag1}) for $N=1$ yields the standard continuity-like equation\footnote{This type of chiral symmetry is broken even for a ``massless'', yet \emph{causal} neutrino due to $M \rightarrow - i\, \varepsilon$!} $\partial_\mu [ \, \overline{\psi^c} (x) \; \gamma^{\,\mu} \, \gamma_5  \psi (x) \, ] \propto M$. The first (broken) chiral symmetry related to the current $\overline{\psi} (x) \, \gamma^{\,\mu} \, \gamma_5  \psi (x)$ mixes causal and anticausal fields, while the second (broken) chiral symmetry related to the current $\overline{\psi^c} (x) \, \gamma^{\,\mu} \, \gamma_5  \psi (x)$ mixes only causal fields or anticausal fields. HQT can't distinguish between the two chiral currents and runs therefore into anomalies yielding different results depending on the choice of regularization \cite{zinn1}. The (anti)causal neutrino Lagrangian (Eq.\ (\ref{diraclag1}) for $N=1$) can be studied, too, by decomposing the the neutrino fields into their chiral components \footnote{$\psi(x) = \psi_R(x) + \psi_L(x)$ with $\psi_R(x):=P_R \, \psi(x)$, $\psi_L(x):=P_L \, \psi(x)$ and $P_R:=(1+\gamma_5)/2$, $P_L:=(1-\gamma_5)/2$.}, i.e.: 
\begin{eqnarray} \lefteqn{{\cal L} (x) =} \nonumber \\
 & = & \frac{1}{2} \Big\{ 
\overline{\psi_L^c} (x) \frac{i}{2} \!\! \stackrel{\;\,\leftrightarrow}{\not\!\partial} \!\! \psi_R (x) + 
\overline{\psi^c_R} (x)  \frac{i}{2} \!\! \stackrel{\;\,\leftrightarrow}{\not\!\partial} \!\! \psi_L (x)  - M \Big(
\overline{\psi_R^c} (x)  \, \psi_R (x) + 
\overline{\psi^c_L} (x)  \, \psi_L (x) \Big)
\Big\} + \mbox{h.c.} \nonumber \\
 & = & \frac{1}{2} \Big\{ 
\overline{\chi_+^c} (x) \frac{1}{2} \!\! \stackrel{\;\,\leftrightarrow}{\not\!\partial} \!\! \chi_+ (x) - 
\overline{\chi^c_-} (x)  \frac{1}{2} \!\! \stackrel{\;\,\leftrightarrow}{\not\!\partial} \!\! \chi_- (x)  -  M \Big(
\overline{\chi_+^c} (x)  \chi_- (x) +  
\overline{\chi^c_-} (x)  \chi_+ (x) \Big)
\Big\} + \mbox{h.c.} \nonumber 
\end{eqnarray}
Note that ${\cal L} (x)$ in terms of the new fields $\chi_\pm (x):= (\psi_R(x) \pm i \psi_L(x))/\sqrt{2}$ is invariant under $\chi_\pm (x) \rightarrow \exp (\pm \,i\,\gamma_5\, \alpha) \, \chi_\pm (x)$ even for \emph{arbitrary complex Fermion mass} $M$!
Further aspects (quantization, Wick's theorem, CPT, $\dots$) of the presented (anti)causal QT are either discussed in Refs.\ \cite{kleefeld1,kleefeld2,kleefeld3,kleefeld4} or will be published elsewhere (e.g. Ref.\ \cite{kleefeld5}).
\section{QCD and the Quark-Level Linear Sigma Model}

As an interesting application of the formalism described above we want to illustrate an attractive relation between the Lagrangian of QCD \cite{qcd1} and the unbroken Lagrangian of the QLLSM \cite{delbourgo1} including vector mesons\footnote{It has to be stated that the presented ``derivation'' is hardly possible in a HQT.}.
We start with the (anti)causal QCD Lagrangian in $R_\alpha$ gauge \cite{jun1,hoo1} for (anti)quarks ($q_\pm$) and gluons ($B$) with respective complex masses $M_q$ and $M_B$ \footnote{For ``massless'' quarks and gluons one has to perform the replacement $M \rightarrow - \, i \,\varepsilon$.}${}^{,}$\footnote{In the following we will use the indices $c$, $S$, $F$ for ``colour'', ``spin'' and ``flavour'', respectively. 

\makebox[0.5mm]{}E.g. $[1]_c$, $[1]_S$ and $[1]_F$ are the unit matrices in colour ($c$), spin ($S$) and flavour ($F$) space.}${}^{,}$\footnote{Be reminded that $D_\mu := \partial_\mu - i \, g \,B_\mu (x)$, $B_\mu (x) := B^a_\mu (x) \; [\,\lambda^a/2\,]_c$, $B_{\mu\nu} (x) := B^{\,a}_{\mu\nu} (x) \; [\,\lambda^a/2\,]_c = \frac{i}{g} \, [D_\mu , D_\nu   ]$, $C (x) := C^{\,a} (x) \; [\,\lambda^a/2\,]_c$ and $\bar{C} (x) := \bar{C}^{\,a} (x) \; [\,\lambda^a/2\,]_c$, while ``$\mbox{tr}_c$'' indicates a trace and $[\,\lambda^a\,]_c$ ($a=1,\ldots,8$) represent Gell-Mann matrices in colour ($c$) space.}:
\begin{eqnarray} \lefteqn{{\cal L} (x) 
 = \overline{q^{\,\,c}_+} (x) \; \left( \frac{i}{2} \! \stackrel{\leftrightarrow}{\not\!\partial} + \; g \not\!B (x) \, - \, M_q \, \right) \; q_- (x) } \nonumber \\
 & - & \frac{1}{2} \; \mbox{tr}_c \Big[ B_{\mu\nu} (x) \, B^{\,\mu\nu} (x) \Big] + \; M_B^2 \;\, \mbox{tr}_c \Big[ \, B_\mu (x) \, B^{\,\mu} (x) \, \Big] \; - \;  \frac{1}{\alpha} \;\, \mbox{tr}_c \Big[ \, \Big(\partial_\mu \, B^{\,\mu} (x)\Big)^2 \, \Big] \nonumber \\
 & - &  2\; \mbox{tr}_c \Big[ \, \Big(\partial_{\mu} \, \bar{C} (x)\,\Big) \; \left( \, \frac{\alpha\, M_B^2}{\partial^2} \, + 1 \, \right) \, \partial^{\,\mu} \, C (x)  + \, i\, g \; \Big\{ \, \Big(\partial_{\mu} \, \bar{C} (x)\,\Big) \, , \, C (x) \, \Big\}  \; B^{\,\mu} (x) \, \Big] + \mbox{h.c.} \nonumber 
\end{eqnarray}
For later convenience we  insert unit matrices in spin and flavour space, i.e.:
\begin{eqnarray} \lefteqn{{\cal L} (x) = \overline{q^{\,c}_+} (x) \; \left( \frac{i}{2} \! \stackrel{\leftrightarrow}{\not\!\partial} + \; g \not\!B (x) \; [\,1\,]_F \, - \, M_q \, \right)  \; q_- (x) } \nonumber \\
 & + & \frac{1}{4\,N_F} \;\, \mbox{tr} \Big[ -\, \frac{1}{2} \; B_{\mu\nu} (x) \; [\,1\,]_S \; [\,1\,]_F\; B^{\,\mu\nu} (x) \; [\,1\,]_S \; [\,1\,]_F \nonumber \\
 & &  + \; M_B^2 \; B_\mu (x) \; [\,1\,]_S \; [\,1\,]_F \; B^{\,\mu} (x) \; [\,1\,]_S \; [\,1\,]_F  - \frac{1}{\alpha}\; \Big(\partial_\mu \, B^{\,\mu} (x) \; [\,1\,]_S \; [\,1\,]_F \Big)^2 \nonumber \\
 & &  -\; 2\; \Big( \, \Big(\partial_{\mu} \, \bar{C} (x) \; [\,1\,]_S \; [\,1\,]_F \,\Big) \; \left( \, \frac{\alpha \; M_B^2}{\partial^2} \, + 1 \, \right) \, \partial^{\,\mu} \, C (x) \; [\,1\,]_S \; [\,1\,]_F \nonumber \\
 & & 
+ \, i\, g \; \Big\{ \, \Big(\partial_{\mu} \, \bar{C} (x) \; [\,1\,]_S \; [\,1\,]_F\,\Big) \, , \, C (x) \; [\,1\,]_S \; [\,1\,]_F \, \Big\}  \; B^{\,\mu} (x) \; [\,1\,]_S \; [\,1\,]_F \, \Big) \, \Big] + \mbox{h.c.}\, , \quad \label{qcdlag1}
\end{eqnarray}
while ``tr'' indicates the trace in colour, spin and flavour space, i.e. ``$\mbox{tr}_{c,S,F}$''. 
For the following arguments it seems imperative to use Feynman gauge ($\alpha \rightarrow 1$) for which the gluon ($i\, D^{\, ab}_{\mu\nu} (k)$) and ghost ($i\, \Delta^{\, ab} (k)$) propagators take their simplest form, i.e.\ $i\, D^{\, ab}_{\mu\nu} (k) \rightarrow - i \, \delta^{\,ab} \;
g_{\mu\nu} /(k^2 - M_B^2)$ and $i\, \Delta^{\, ab} (k) \rightarrow  - i \, \delta^{\,ab} /(k^2 - \, M_B^2)$.
We know that at high energies --- in the perturbative regime of QCD --- quark--quark scattering is predominantly determined by one-gluon exchange (OGE) whose spin structure is --- in Feynman gauge --- described by $[\, \gamma_{\,\mu} \, ]^{(1)} \; [\, \gamma^{\,\mu} \, ]^{(2)}$ being subject to the Fierz identity \cite{alk1} 
\[ [\,\gamma_\mu\,]_{\,ij} [\,\gamma^{\,\mu}\,]_{\,k\ell} = [\,1\,]_{\,i\ell} [\,1\,]_{\,kj} + [\,i \, \gamma_5 \,]_{\,i\ell} [\,i\,\gamma_5\,]_{\,kj} - \frac{1}{2} \; \Big( \,  [\,\gamma_\mu\,]_{\,i\ell} [\,\gamma^{\,\mu}\,]_{\,kj} + [\,\gamma_\mu\gamma_5 \,]_{\,i\ell} [\,\gamma^{\,\mu}\gamma_5 \,]_{\,kj} \, \Big)\, . \]
Inspection of these Fierz identities suggests that a $t$-channel one-gluon OGE can be replaced by the simultaneous $u$-channel exchange ($j \leftrightarrow \ell$) of scalar ($S$), pseudo-scalar ($P$), vector ($V$) and axial-vector ($Y$) fields, while the $u$-channel OGE may be replaced by the respective simultaneous $t$-channel exchange of $S$, $P$, $V$ and $Y$ fields. Of course, this idea can't be considered isolated from the respective dynamics in colour and flavour space \footnote{This statement applies to QCD. In a similiar (feasible) consideration for QED in the perturbative (low energy) regime the Fierz identities in colour and flavour space are absent. The results are expected to be similar to the observations in \cite{haymaker1}.}. The relevant Fierz identity in colour space is:
\[ {[\,\lambda^{a}/2\,]_{\,ij} \; [\,\lambda^{a}/2\,]_{\,k\ell} }
 =  \frac{1}{2} \, \Big( 1 - \, \frac{1}{N^2_c} \,\Big) \; [\,1\,]_{\,i\ell} \; [\,1\,]_{\,kj} - \, \frac{1}{N_c} \; [\,\lambda^{a}/2\,]_{\,i\ell} \; [\,\lambda^{a}/2\,]_{\,kj} \; , \]
while in flavour space we have $ {[\,1\,]_{\,ij} \; [\,1\,]_{\,k\ell}} 
 = \frac{1}{N_F} \;  [\,1\,]_{\,i\ell} \; [\,1\,]_{\,kj}  \; + \; 2\; 
{ [\,\lambda^{a}/2\,]_{\,i\ell} \; [\,\lambda^{a}/2\,]_{\,kj}}$.
The Fierz identity in flavour space seems to support the idea that the exchanged $S$, $P$, $V$ and $Y$ particles underlying the $t$-channel OGE are e.g.\ flavour nonets (for $N_F=3$), while the Fierz identity in colour space indicates that in the large $N_c$ limit ($N_c \rightarrow \infty$) these $S$, $P$, $V$ and $Y$ fields may be considered to be colour singlets. Hence one might tend to call these exchanged $S$, $P$, $V$ and $Y$ fields either ``mesons'' or ``glue balls''. Defining $[\,\lambda^{0}/2\,]_F := [\,1\,]_F /\sqrt{2\,N_F}$ and keeping in mind that in quark-quark scattering always $t$- and $u$-channel OGE graphs contribute both --- due to the Pauli exclusion principle with opposite sign --- we may claim the validity of the following Fierz replacement in the product of spin, flavour and colour space:
\begin{eqnarray}  \lefteqn{\Big( [\, \gamma_\mu \, ]^{(1)} \; [\, \gamma^{\,\mu} \, ]^{(2)} \Big) \; 
\Big( [\, \lambda^a/2 \, ]_c^{(1)} \; [\, \lambda^a/2 \, ]_c^{(2)} \Big) \; 
\Big( [\, 1 \, ]_F^{(1)} \; [\, 1 \, ]_F^{(2)} \Big) } \nonumber \\
 & \longrightarrow & 
- \Big( [\,1\,]^{(1)} [\,1\,]^{(2)} + [\,i \, \gamma_5 \,]^{(1)} [\,i\,\gamma_5\,]^{(2)}  -\frac{1}{2} \; \Big\{ \,  [\,\gamma_\mu\,]^{(1)} [\,\gamma^{\,\mu}\,]^{(2)} + [\,\gamma_\mu\gamma_5 \,]^{(1)} [\,\gamma^{\,\mu}\gamma_5 \,]^{(2)}  \Big\} \Big) \nonumber \\
 & & \quad \Big( \frac{1}{2} \, \Big( 1 - \, \frac{1}{N^2_c} \, \Big) \; [\,1\,]_c^{(1)} [\,1\,]_c^{(2)} - \, \frac{1}{N_c} \; [\,\lambda^{a}/2\,]_c^{(1)} [\,\lambda^{a}/2\,]_c^{(2)} \Big) \nonumber \\
 & & \quad \Big( \, \frac{1}{N_F} \;  [\,1\,]_F^{(1)} [\,1\,]_F^{(2)}  \; + \; 2\;\, [\,\lambda^{a}/2\,]_F^{(1)} [\,\lambda^{a}/2\,]_F^{(2)} \Big) \nonumber \\
 & \stackrel{N_c \rightarrow \infty}{\longrightarrow} & 
- \Big( [\,1\,]^{(1)} [\,1\,]^{(2)} + [\,i \, \gamma_5 \,]^{(1)} [\,i\,\gamma_5\,]^{(2)}  -\frac{1}{2} \; \Big\{ \,  [\,\gamma_\mu\,]^{(1)} [\,\gamma^{\,\mu}\,]^{(2)} + [\,\gamma_\mu\gamma_5 \,]^{(1)} [\,\gamma^{\,\mu}\gamma_5 \,]^{(2)} \Big\} \Big) \nonumber \\
 & & \quad \Big( \,  [\,1\,]_c^{(1)} [\,1\,]_c^{(2)} \Big)  \; \Big(  [\,\lambda^{0}/2\,]_F^{(1)} [\,\lambda^{0}/2\,]_F^{(2)} \; + \; [\,\lambda^{a}/2\,]_F^{(1)} [\,\lambda^{a}/2\,]_F^{(2)} \Big) \, . \label{fierep1}
\end{eqnarray}
After defining $N_F\times N_F$ ``meson'' \footnote{In the following we will use for simplicity the notatation ``meson'' rather than ``glue ball''.} field matrices in flavour space ($\sum := \sum_{a=0}^{N^2_F-1}$), i.e.\ $S (x) := \sqrt{2} \;\sum \; \sigma_a (x) \; [\,\lambda^{a}/2\,]_F$ (``scalar''), $P (x) := \sqrt{2} \;\sum \; \eta_a (x) \; [\,\lambda^{a}/2\,]_F$ (``pseudo--scalar''), $V^{\mu} (x) :=  \sqrt{2} \;\sum \; \omega^{\,\mu}_a (x) \; [\,\lambda^{a}/2\,]_F$ (``vector'') and $Y^{\mu} (x) := \sqrt{2} \;\sum \; \upsilon^{\,\mu}_a (x) \; [\,\lambda^{a}/2\,]_F$ (``axial--vector'') 
we may translate Fierz replacement Eq.\ (\ref{fierep1}) for $N_c \rightarrow \infty$ into a respective replacement prescription for the OGE propagator in terms of meson propagators:
\begin{eqnarray} \lefteqn{[\, \gamma^{\,\mu} \, ]^{(1)} \; [\, \gamma^{\,\nu} \, ]^{(2)} \;  \left< 0\right|T\Big[ \,B^{(1)}_\mu (x)\; B^{(2)}_\nu (y) \,\Big]\left|0\right> \;\, \Big( \,  [\,1\,]_F^{(1)} \; [\,1\,]_F^{(2)} \Big) \quad \stackrel{N_c \rightarrow \infty}{\longrightarrow}
} \nonumber \\
 &  & \frac{1}{2} \; \Big( \; [\,1\,]^{(1)} [\,1\,]^{(2)} \; Z^{\,-1}_s \, \left<0\right|T\Big[ \,S^{(1)} (x)\; S^{(2)} (y) \,\Big]\left|0\right> \nonumber \\ 
 & & \quad + [\,i \, \gamma_5 \,]^{(1)} [\,i\,\gamma_5\,]^{(2)} \;\; Z^{\,-1}_p \, \left<0\right|T\Big[ \,P^{(1)} (x)\; P^{(2)} (y) \,\Big]\left|0\right> \nonumber \\ 
 & & \quad +  \frac{1}{2} \Big\{ [\,\gamma^{\,\mu}\,]^{(1)} [\,\gamma^{\,\nu}\,]^{(2)} \; Z^{\,-1}_v \, \left<0\right|T\Big[ \,V^{(1)}_\mu (x)\; V^{(2)}_\nu (y) \,\Big]\left|0\right> \nonumber \\
 & & \quad \quad \;\; + [\,\gamma^{\mu}\gamma_5 \,]^{(1)} [\,\gamma^{\,\nu}\gamma_5 \,]^{(2)} \; Z^{\,-1}_y \left<0\right|T\Big[ \,Y^{(1)}_\mu (x)\, Y^{(2)}_\nu (y) \,\Big]\left|0\right> \Big\} \Big) \Big( \,  [\,1\,]_c^{(1)} [\,1\,]_c^{(2)} \Big) \, . \quad\; \label{prorep1} \end{eqnarray}
The renormalization constants $Z_s$, $Z_p$, $Z_v$ and $Z_y$ will be determined later. Treating each element of the meson field matrices $S(x)$, $P(x)$, $V^\mu(x)$ and $Y^\mu(x)$ as an independent field the propagator replacement Eq.\ (\ref{prorep1}) will be achieved by replacing in the quark-gluon vertex of the QCD-Lagrangian ($g\;\, \overline{q^{\,\,c}_+} (x) \not\!B (x) \, [\,1\,]_F \; q_- (x)$) the gluon field according to\footnote{Remember that $\gamma_\mu \gamma^\mu = 4 \; [1]_S$.}:
\begin{eqnarray} \not\!B (x) \;\, [\,1\,]_F & \stackrel{N_c \rightarrow \infty}{\longrightarrow} & \frac{1}{\sqrt{2}} \; \gamma_\mu \; \Big( \, \frac{1}{4} \; \Big( \, \frac{c_s}{\sqrt{Z_s}} \; \gamma^{\, \mu} \; S(x) + \frac{c_p}{\sqrt{Z_p}} \; i\,\gamma^{\, \mu} \; \gamma_5 \, P(x) \, \Big) \nonumber \\
 & & \qquad\quad + \frac{1}{\sqrt{2}} \; \Big( \, \frac{c_v}{\sqrt{Z_v}} \; V^{\,\mu} (x) \; [\,1\,]_S + \frac{c_y}{\sqrt{Z_y}} \; Y^{\,\mu} (x) \, \gamma_5 \, \Big) \, \Big) \; [\,1\,]_c \quad \label{brep1}
\end{eqnarray}
with $c_s,c_p,c_v,c_y \in\{+ 1,-1\}$. Inspection of Eq.\ (\ref{brep1}) and $\not\!B (x) = \gamma_\mu \, B^\mu (x)$ suggests the following large $N_c$ replacement of the gluon fields in terms of $S$, $P$, $V$ and $Y$ mesons\footnote{It should be noted that the suggested replacement is not completely unique. Nevertheless it is strongly suggested by the requirement of renormalizability of the resulting Lagrangian and the structure of Lagrangians obtained in the context of extended Nambu-Jona-Lasinio models.}:
\begin{eqnarray}  B^\mu (x) \;\, [\,1\,]_S\, [\,1\,]_F & \stackrel{N_c \rightarrow \infty}{\longrightarrow} & \frac{1}{\sqrt{2}} \; \Big( \, \frac{1}{4} \; \Big( \, \frac{c_s}{\sqrt{Z_s}} \; \gamma^{\, \mu} \; S(x) + \frac{c_p}{\sqrt{Z_p}} \; i\,\gamma^{\, \mu} \; \gamma_5 \, P(x) \, \Big) \nonumber \\
 & & \quad + \frac{1}{\sqrt{2}} \; \Big( \, \frac{c_v}{\sqrt{Z_v}} \; V^{\,\mu} (x) \; [\,1\,]_S + \frac{c_y}{\sqrt{Z_y}} \; Y^{\,\mu} (x) \, \gamma_5 \, \Big) \, \Big) \; [\,1\,]_c\, . \quad \label{bbrep1}
\end{eqnarray}
Using Eq.\ (\ref{bbrep1}) it is also straight forward to replace the gluon field-strength tensor $B_{\mu\nu} (x)$ keeping in mind that $B_{\mu\nu} (x) \; [\,1\,]_S \; [\,1\,]_F = \frac{i}{g} \,[ \, D_\mu \; [\,1\,]_S \; [\,1\,]_F \; , \; D_\nu  \; [\,1\,]_S \; [\,1\,]_F ]  = [ \,\partial_\mu \, , \, B_\nu (x) \; [\,1\,]_S [\,1\,]_F \, ] - [ \, \partial_\nu \, , \, B_\mu (x) \; [\,1\,]_S [\,1\,]_F \, ] - i \, g \, [ \, B_\mu (x) \; [\,1\,]_S [\,1\,]_F \, , \, B_\nu (x) \; [\,1\,]_S [\,1\,]_F \,]$.
In order to guarantee the renormalizability of the resulting Lagrangian the Grassmann ghosts $C (x)$ and $\bar{C} (x)$ may by decomposed for the $V$ and $Y$ fields into $C_v (x) \; c_v/\sqrt{Z_v}$, $\bar{C}_v (x)\; c_v/\sqrt{Z_v}$, $C_y (x)\; c_y/\sqrt{Z_y}$ and $\bar{C}_y (x)\; c_y/\sqrt{Z_y}$ with  $C_v (x) = \sum C_v^{\,a} (x) \; [\,\lambda^a/2\,]_F$, $C_y (x) = \sum C_y^{\,a} (x) \; [\,\lambda^a/2\,]_F$, $\bar{C}_v (x) = \sum \bar{C}_v^{\,a} (x) \; [\,\lambda^a/2\,]_F$, $\bar{C}_y (x) = \sum \bar{C}_y^{\,a} (x) \; [\,\lambda^a/2\,]_F$ and $\sum = \sum^{N^2_F-1}_{a=0}$. The scalar and pseudo-scalar fields renormalize each other in the sense of the LSM, if they form a chiral circle. After application of the replacements Eq.\ (\ref{brep1}) and (\ref{bbrep1}) to the QCD Lagrangian Eq.\ (\ref{qcdlag1}) we may choose $Z_s$, $Z_p$, $Z_v$ and $Z_y$ such that the kinetic terms of the new $S$, $P$, $V$ and $Y$ fields take the standard form. This is achieved by:
\begin{equation}  \frac{c_\ell}{\sqrt{Z_\ell}} \; = \; \frac{4\, i \, s_\ell}{\sqrt{3+\frac{1}{\alpha}}} \; \sqrt{\frac{N_F}{N_c}} \quad (\ell\in \{s,p\}) \;\; , \quad  
 \frac{c_{\ell^\prime}}{\sqrt{Z_{\ell^\prime}}} \; = \; \sqrt{2} \; s_{\ell^\prime} \; \sqrt{\frac{N_F}{N_c}} \quad (\ell^\prime \in \{v,y\}) \; , 
  \label{renorm1}
\end{equation}
 with $s_s,s_p,s_v,s_y \in\{+ 1,-1\}$. As an overall result of our replacements we obtain in Feynman gauge ($\alpha \rightarrow 1$) the following renormalizable Lagrangian of a QLLSM including (axial)vector-mesons, which is expected to describe the properties of QCD at high energies\footnote{This statement should be understood in the sense of conclusions drawn in Ref.\ \cite{dass1} by N.D.\ Hari Dass and V.\ Soni. In their paper they discuss ``non-Abelian gauge theories with fermions and scalars that nevertheless possess asymptotic freedom''. They claim that ``even the possibility that'' such a theory (being hardly distinguishable from the QLLSM including vector-mesons) ``is indistinguishable from QCD could not be ruled out''. Their analysis shows that ``for deep inelastic scattering the leading behavior of this theory in the ultraviolet is identical to that of QCD''.}${}^{,}$\footnote{We want to mention that for the QCD anomalous term we find the replacement  $\mbox{tr}_{c} [ \, B^{\,\mu\nu} (x) \, \tilde{B}_{\mu\nu} (x) \, ] \stackrel{N_c \rightarrow \infty}{\longrightarrow}\frac{1}{4} \; \mbox{tr}_{F} [ \, V_+^{\,\mu\nu} (x) \, \tilde{V}_{+\,\mu\nu} (x) \, ] + \frac{1}{4} \; \mbox{tr}_{F} [ \, V_-^{\,\mu\nu} (x) \, \tilde{V}_{-\,\mu\nu} (x) \, ]$ (affecting only (axial--)vector mesons). In order to obtain anomalous structures like the 't~Hooft determinant (see e.g.\ Ref.\ \cite{hoo4}) involving also (pseudo-)scalars we suggest to apply our replacement strategy to vanishing terms like e.g.\ $\mbox{tr}_c [ \, B^{\,\mu\nu} (x) \; B_{\mu\nu} (x) \; \gamma_5 \, ]$.}:
\begin{eqnarray} \lefteqn{{\cal L} (x) \; \stackrel{N_c \rightarrow \infty}{\longrightarrow}}  \nonumber \\
 & \rightarrow & 
\overline{q^{\,c}_+} (x) \; \left( \frac{i}{2} \! \stackrel{\leftrightarrow}{\not\!\partial} \, - \, M_q  \, \right) \; q_- (x) \nonumber \\
 & + & \overline{q^{\,c}_+} (x) \, g \sqrt{\frac{N_F}{N_c}} \Big( \sqrt{2}\; i  \Big( s_s \, S(x) + s_p \, i\,\gamma_5 \, P(x) \Big) 
+  \frac{1}{\sqrt{2}} \, \Big( s_v \, \not\!V (x) + s_y \, \not Y (x) \, \gamma_5\Big) \, \Big) q_- (x)  \nonumber \\
 & + & \frac{1}{8} \; \mbox{tr}_{F} \Big[ \Big( \partial^\mu \, S(x) \Big)  \Big( \partial_\mu \, S(x) \Big)  \Big] - \, \frac{1}{2} \;   M_B^2 \; \mbox{tr}_{F} \Big[  \Big( S(x) \Big)^2 \; \Big]  \nonumber \\
 & + &  \frac{1}{8} \; \mbox{tr}_{F} \Big[ \Big( \partial^\mu \, P(x) \Big)  \Big( \partial_\mu \, P(x) \Big)  \Big] - \, \frac{1}{2} \;  M_B^2 \; \mbox{tr}_{F} \Big[  \Big( P(x) \Big)^2 \; \Big] \nonumber \\
 & + & \frac{3}{8} \; 
\mbox{tr}_{F} \Big[ \Big( \partial^\mu \, S(x)   
-\,i\, g \; \sqrt{\frac{N_F}{2\,N_c}} 
\Big( s_v \; [ \, V^\mu (x)\; , \; S(x)\, ] - i \, s_s \,s_p \, s_y \; \{ \, Y^{\, \mu}(x)\; , \; P(x) \, \} \Big)
\Big)^2 \;   \Big] \nonumber \\ 
 & + & \frac{3}{8} \; 
\mbox{tr}_{F} \Big[ \Big( \partial^\mu \, P(x)   
-\,i\, g \; \sqrt{\frac{N_F}{2\,N_c}} 
\Big( s_v \; [ \, V^\mu (x)\; , \; P(x)\, ] + i \, s_s \,s_p \, s_y \; \{ \, Y^{\, \mu}(x)\; , \; S(x) \, \} \Big)
\Big)^2 \;  \Big] \nonumber \\ 
 & - & \frac{1}{8} \; \mbox{tr}_{F} \Big[ \, \Big(V_+^{\,\mu\nu} (x)\Big)^2 \, \Big] + \frac{1}{2} \; M_B^2 \; \mbox{tr}_{F} \Big[\; \Big( V^{\,\mu} (x)\Big)^2 \, \Big]  - \; \frac{1}{2} \; \mbox{tr}_{F} \Big[\; \Big( \partial^\mu \, V_\mu (x) \, \Big)^2  \; \Big] \nonumber \\
 & - & \frac{1}{8} \; \mbox{tr}_{F} \Big[ \, \Big(V_-^{\,\mu\nu} (x)\Big)^2 \,  \Big]  +  \frac{1}{2} \;  M_B^2 \; \mbox{tr}_{F} \Big[ \; \Big(Y^\mu (x)\Big)^2 \; \Big]  - \; \frac{1}{2} \; \mbox{tr}_{F} \Big[\; \Big( \partial^\mu \, Y_\mu (x) \, \Big)^2  \; \Big] \nonumber \\
 & + & \frac{3}{8} \;
\left(-\,i\, g \; \sqrt{\frac{N_F}{N_c}} \; \right)^2  \; \mbox{tr}_{F} \Big[  \Big( \Big( S(x)  + \, i \, P(x) \Big) \Big( S(x)  - \, i \, P(x) \Big) \Big)^2  \,  \Big] \nonumber \\
 & - &  \mbox{tr}_F \Big[ \, \Big(\partial_{\mu} \, \bar{C}_v (x)\,\Big) \; \Big( \, \frac{M_B^2}{\partial^2} \, + 1 \, \Big) \partial^{\,\mu} C_v (x)  +  i\, g \sqrt{\frac{N_F}{2\, N_c}} \Big\{ \Big(\partial_{\mu} \, \bar{C}_v (x)\,\Big) \, , \, C_v (x) \Big\}  \, V^{\,\mu} (x) \Big] \nonumber \\
 & - & \mbox{tr}_F \Big[ \, \Big(\partial_{\mu} \, \bar{C}_y (x)\,\Big) \; \Big( \, \frac{M_B^2}{\partial^2} \, + 1 \, \Big) \partial^{\,\mu} C_y (x)  +  i\, g \sqrt{\frac{N_F}{2\, N_c}} \Big\{ \Big(\partial_{\mu} \, \bar{C}_y (x)\,\Big) \, , \, C_y (x) \Big\}  \, Y^{\,\mu} (x) \Big] \nonumber \\
 & + & \mbox{h.c.} \label{qllsmlag1}
\end{eqnarray} 
with $V_\pm^{\,\mu\nu} (x) :=  [\,{\cal D}_\pm^\mu \,,\,{\cal D}_\pm^\nu\, ] / \left( - i \, g \; \sqrt{N_F/(2\,N_c)} \; \right)$ being defined by the covariant derivatives ${\cal D}_\pm^\mu := \partial^\mu - i \, g \; \sqrt{N_F/(2\,N_c)} \,( \, s_v  \, V^{\,\mu} (x) \pm s_y  \, Y^{\,\mu} (x) \, )$. In spite of limited space a discussion of Eq.\ (\ref{qllsmlag1}) is imperative. Apart from slight, yet significant differences the Lagrangian is very similar to the Lagrangian of the ``traditional'' QLLSM (see e.g.\ Refs. \cite{delbourgo1,scadron1,beveren1} and references therein). The $V$ and $Y$ mesons couple to the $S$ and $P$ mesons in a similar manner as in an 1-loop effective action of the Gauged LSM in Ref.\ \cite{alk1} or the Extended Chiral Quark Model in Ref.\ \cite{andrianov1}  obtained by a tedious Bosonization. The individual interaction terms in Eq.\ (\ref{qllsmlag1}) follow strictly the weighting rules of the $1/N_c$ expansion (see \cite{hoo2} and Chapter 8 in \cite{qcd1} including references). A characteristic new property of Eq.\ (\ref{qllsmlag1}) is that the couplings of the scalar and pseudo-scalar fields to (anti)quarks contain an extra factor $\, i\,$ (= imaginary unit)\footnote{This complex phase being related to the choice of $Z_s$ and $Z_p$ in Eq.\ (\ref{renorm1}) maps the negative spatial components of the $(+,-,-,-)$ metric into its positive time-like component yielding negative relative signs between the mass or kinetic terms of (axial-)vector fields relative to (pseudo-)scalar fields. In the language of Ref.\ \cite{haymaker1} it would map a ``wrong sign mass term'' into a right sign mass term or vice-versa.}. It appears as a surprise that the quasi-Hermitian QCD Lagrangian has been translated at high energies into a seemingly non-Hermitian Lagrangian with the same properties. Non-Hermitian QT allows now to relate consistently the high-energy Lagrangian to a low-energy Lagrangian by performing a Higgs-mechanism (HM) with a complex-valued scalar condensate $\left< S\right>$ yielding at low energies complex selfenergies of mesons and (anti)quarks and complex coupling constants without getting in conflict with unitarity, causality, renormalizability and Lorentz covariance. E.g. a real-valued scalar condensate $\left< S\right>$ would yield --- due to the additional phasefactor in the coupling of scalars to (anti)quarks ---  purely imaginary selfenergy for (anti)quarks\footnote{Confinement would therefore just be a notation for highly unstable (or composite), i.e. very short lived, (anti)quarks decaying at low energies very fast into some asymptotic colourless meson states with quasi-real selfenergies}. There is also the scenario of (close to) purely imaginary scalar condensates $\left< S\right>$ yielding on one hand quasi-real (anti)quark selfenergies, on the other hand ``vacuum replica'' \cite{bicudo1} or ``vacuum instability'' like structures \cite{rembiesa1} in the ``composite field-space'' \footnote{In the discussion of the ``Composite-Field Effective Potential'' \cite{haymaker1,rembiesa1} the role of the scalar fields is represented by so called ``composite fields''. These considerations find their origin in a stability and convergence analysis of the parameter space of the LSM which may be found in Ref. \cite{okubo1}.}${}^{,}$\footnote{We warmly like to thank J.E.F.T.\ Ribeiro for illustrating --- in private communication --- properties of ``Vacuum Replicas'' \cite{bicudo1} and drawing our attention to related work of P.\ Rembiesa (e.g.\ Ref.\ \cite{rembiesa1})!}.
A further advantage of the derived QLLSM Lagrangian is that one gets a very clear understanding how the electromagnetic interaction couples to quarks and mesons. Photons couple in the QCD Lagrangian only to (anti)quarks, not to gluons, which yields that they also don't couple at high energies to the new $S$, $P$, $V$ and $Y$ fields. A complex Renormalization Group Transformation or --- equivalently --- a complex-valued HM relating Eq.\ (\ref{qllsmlag1}) to the  low energy version of the QLLSM Lagrangian may mix the photon fields with the vector-meson fields such that the photons couple solely to vector-mesons and not to quarks\footnote{This scenario is commonly called Vector-Meson Dominance (VMD).}. The discussed implementation of QED in QCD makes clear why one faces  double-counting ambiguities \cite{bramon1} at low energies, whenever one tries to couple photons to both, (anti)quarks \emph{and} mesons. 
\begin{theacknowledgments}
We particularly want to thank George Rupp, Eef van Beveren and Jun-Chen Su for very valuable remarks. 
A comprehensive form of the mainly unpublished results has been presented for the first time in a seminar (5.2.2002) and lecture course (21.3.-11.4.2002) in the CFIF, Lisbon. 
This work has been supported by the 
{\em Funda\c{c}\~{a}o para a Ci\^{e}ncia e a Tecnologia} \/(FCT) of the {\em Minist\'{e}rio da Ci\^{e}ncia e da Tecnologia (e do Ensinio Superior)} \/of Portugal, under Grant no.\ PRAXIS
XXI/BPD/20186/99 and SFRH/BDP/9480/2002.
\end{theacknowledgments}




\begin{thebibliography}{xx}
\bibitem{araki1} H. Araki, \emph{Mathematical theory of quantum fields}, \copyright 1999 by Oxford University Press;
Lecture Notes by F. Strocci, \emph{General properties of QFT} (Lecture Notes in Physics --- Vol.\ 51), \copyright 1993 by World Scientific; R. Haag, \emph{Local Quantum Physics} (2nd ed.), \copyright 1992, 1996 by Springer-Verlag; R.\ Jost, \emph{The general theory of quantized fields}, Providence, R.I.: Am.\ Phys.\ Soc., 1965; R.F.\ Streater, A.S.\ Wightman, \emph{PCT, spin and statistics, and all that}, Benjamin, New York, 1964; N.N.\ Bogoliubov, D.V.\ Shirkov, \emph{Introduction to the Theory of Quantized Fields}, \copyright 1980 by John-Wiley.
\bibitem{mueckenheim1} W.\ M\"uckenheim$\;$ et al., \emph{Phys.\ Rep.} \textbf{\bf 133} (1986) 337. 
\bibitem{afriat1} A.\ Afriat, F.\ Selleri, 
                      \emph{The Einstein, Podolski, and Rosen Paradox in Atomic, Nuclear and Particle Physics}, 
                      \copyright 1999 by Plenum Press, New York. 
\bibitem{greiner1} W.\ Greiner, B.\ M\"uller, J.\ Rafelski, 
                      \emph{Quantum Electrodynamics of Strong Fields}, 
                      \copyright 1985 by Springer-Verlag Berlin Heidelberg; W.\ Greiner, 
                      \emph{Relativistic Quantum Mechanics --- Wave Equations}, 
                      \copyright 1997 by Springer-Verlag Berlin Heidelberg; S.D. Joglekar, hep-th/0106264. 
\bibitem{lee1} T.D.\ Lee, \emph{Phys.\ Rev.} \textbf{\bf 95} (1954) 1329. 
\bibitem{fradkin1} E.S.\ Fradkin, ``Method of Green's Functions in Quantum Field Theory and in Quantum Statistics'', in
                      \emph{Quantum Field Theory and Hydrodynamics}, edited by D.V.\ Skobel'tsyn, Proc. of the P.N. Lebedev Physics Inst. of the Academy of Sciences of the USSR, Moscow, Vol. 29, 
                      \copyright 1967 by Consultants Bureau, Plenum Publishing Corp., New York (Lib. of Congress Cat.\ Card No. 66-12629), pp. 1--132. 
\bibitem{zinn1} J.\ Zinn-Justin, ``The Regularization Problem and Anomalies in Quantum Field Theory'', in Proc. XVIII Lisbon Autumn School on
                      \emph{Topology of Strongly Correlated Systems} (October 8--13, 2000, CFIF, Lisbon, Portugal), edited by P.\ Bicudo, J.E.\ Ribeiro, P.\ Sacramento, V.\ Seixas, V.\ Vieira, 
                      \copyright 2001 by World Scientific Publishing Co.\ Pte. Ltd.  (ISBN 981-02-4572-6), pp. 141--183. 
\bibitem{benayoun1} M.\ Benayoun, H.B.\ O'Connell, \emph{Eur.\ Phys.\ J.} \textbf{\bf C 22} (2001) 503. 
\bibitem{kniehl1} B.A.\ Kniehl, A.\ Sirlin, \emph{Phys.\ Lett.} \textbf{\bf B 530} (2002) 129.
\bibitem{espriu1} D.\ Espriu, J.\ Manzano, P.\ Talavera, \emph{Phys.\ Rev.} \textbf{\bf D 66} (2002) 076002.
\bibitem{gouvea1} A.\ de Gouv\^{e}a, B.\ Kayser, R.N. Mohapatra, hep-ph/0211394.
\bibitem{delia1} M.\ D'Elia, M.-P.\ Lombardo, hep-lat/0209146, hep-lat/0205022.
\bibitem{dass2} N.D.\ Hari Dass, V.\ Soni, hep-th/0204177. 
\bibitem{kleefeld1} F.\ Kleefeld, E. van Beveren, G. Rupp, \emph{Nucl.\ Phys.} \textbf{\bf A 694} (2001) 470.
\bibitem{kaldass1} H.\ Kaldass, A.\ Bohm, S.\ Wickramasekara, \emph{Int.\ J.\ Mod.\ Phys.} \textbf{\bf A 17} (2002) 3749; R.\ de la Madrid, M.\ Gadella, \emph{Am.\ J.\ Phys.} \textbf{70} (2002) 626 (= quant-ph/0201091);
R.C.\ Fuller, \emph{Phys.\ Rev.} \textbf{\bf 188} (1969) 1649;
A.O.\ Barut, ``Dispersion Relations and Resonance Scattering'', in \emph{Lectures in Theoretical Physics} (Vol. IV), delivered at Summer Inst. Theor. Phys., Univ. Colorado, Boulder, 1961, edited by W.E.\ Brittin et al., 
                      \copyright 1962 by John Wiley \& Sons (Lib. of Congress Cat.\ Card No. 59-13034), pp. 460--523. 
\bibitem{moiseyev1} N.\ Moiseyev, \emph{Phys.\ Rep.} \textbf{\bf 302} (1998) 211. 
\bibitem{nakanishi1} N.\ Nakanishi,  \emph{Prog.\ Theor.\ Phys.} \textbf{\bf 19} (1958) 607.
\bibitem{nakanishi2} N.\ Nakanishi,  \emph{Prog.\ Theor.\ Phys.\ Suppl.} \textbf{\bf 51} (1972) 1.
\bibitem{nakanishi3} N.\ Nakanishi,  \emph{Phys.\ Rev.} \textbf{\bf D 5} (1972) 1968.
\bibitem{stapp1} H.P.\ Stapp,  \emph{Il Nuovo Cimento} \textbf{\bf 23 A} (1974) 357.
\bibitem{regge1} T.\ Regge,  \emph{Il Nuovo Cimento (Serie X)} \textbf{\bf 18} (1960) 947.
\bibitem{heisenberg1} W.\ Heisenberg, ``Einf\"uhrung in die Theorie der Elementarteilchen'', lecture given at University of Munich in 1961,
lecture notes prepared by H.\ Rechenberg, K.\ Lagally (ICTP Trieste library number 539.12 H473), see p.\ 95 ff. 
\bibitem{hittner1} D.\ Hittner,  \emph{Il Nuovo Cimento} \textbf{\bf 15 A} (1973) 401.
\bibitem{casagrande1} F.\ Casagrande, L.A.\ Lugiato,  \emph{Il Nuovo Cimento} \textbf{\bf 17 A} (1973) 464.
\bibitem{rabuffo1} I.\ Rabuffo, G.\ Vitiello,  \emph{Il Nuovo Cimento} \textbf{\bf 44 A} (1978) 401.
\bibitem{tarrach1} R.\ Tarrach, hep-th/9502020.
\bibitem{ahlig1} S.\ Ahlig, R.\ Alkofer, \emph{Annals Phys.} \textbf{\bf 275} (1999) 113. 
\bibitem{sauli1} V.\ \v{S}auli, hep-ph/0211221, hep-ph/0209046. 
\bibitem{maris1} P.\ Maris, \emph{Phys.\ Rev.} \textbf{\bf D 52} (1995) 6087; \emph{Phys.\ Rev.} \textbf{\bf D 50} (1994) 4189.
\bibitem{schreiber1} A.W.\ Schreiber, R.\ Rosenfelder, C.\ Alexandrou, hep-th/0007182; C.\ Alexandrou, R.\ Rosenfelder, A.W.\ Schreiber, \emph{Phys.\ Rev.} \textbf{\bf D 62} (2000) 085009. 
\bibitem{helen1} Helen R.\ Quinn, privat communication (May 7, 2002).
\bibitem{mostafazadeh1} A.\ Mostafazadeh, math-ph/0209018; G.\ Scolarici, L.\ Solombrino, quant-ph/0211161.
\bibitem{dirac1} P.A.M.\ Dirac, \emph{Proc.\ Roy.\ Soc.\ (London)} \textbf{\bf A 180} (1942) 1.
\bibitem{dirac2} P.A.M.\ Dirac, \emph{Comm.\ Dublin Inst.\ Adv.\ Studies, A.}, No.\ 1 (1943).
\bibitem{pauli1} W.\ Pauli, \emph{Rev.\ Mod.\ Phys.} \textbf{\bf 15} (1943) 175; S.N.\ Gupta, \emph{Proc.\ Phys.\ Soc.} \textbf{\bf 63} (1950) 681 and \emph{Proc.\ Phys.\ Soc.} \textbf{\bf 64} (1951) 850; K.\ Bleuler, \emph{Helv.\ Phys.\ Acta} \textbf{\bf 23} (1950) 567; K.\ Bleuler, W.\ Heitler, \emph{Prog.\ Theor.\ Phys.}  \textbf{\bf 5} (1950) 600; J.M.\ Jauch, R.\ Rohrlich, \emph{The theory of photons and electrons} (2nd corr.\ printing 1980), published by Springer-Verlag, \copyright 1955 by J.M.\ Jauch, R.\ Rohrlich, see p.\ 103 ff; E.C.G. Sudarshan, ``Lectures in theoretical physics'', 
lecture given at Madras Univ.\ in 1970/71,
lect.\ notes prepared by M.\ Seetharaman, K.\ Eswaran (available in the ICTP Trieste library).
\bibitem{lehmann1} H.\ Lehmann, \emph{Il Nuovo Cimento} \textbf{\bf 11} (1954) 342; G.\ K\"all\'{e}n, \emph{Helv.\ Phys.\ Acta} \textbf{\bf 25} (1952) 417. 
\bibitem{schwinger1} J.\ Schwinger, \emph{Annals Phys.} \textbf{\bf 9} (1960) 169. 
\bibitem{spitzenberg1} T.\ Spitzenberg, K.\ Schwenzer, H.-J.\ Pirner, \emph{Phys.\ Rev.} \textbf{\bf D 65} (2002) 074017; M.A.\ Stephanov, \emph{Nucl.\ Phys.} \textbf{\bf B} \emph{(Proc.\ Suppl.)} \textbf{\bf 53} (1997) 469.
\bibitem{markushin1} V.E.\ Markushin, R.\ Rosenfelder, A.W.\ Schreiber, \emph{Il Nuovo Cimento} \textbf{\bf 117 B} (2002) 75. 
\bibitem{pais1} A.\ Pais, ``Max Born and the statistical interpretation of QM'', KEK library preprint 198301341.
\bibitem{weldon1} H.A.\ Weldon, \emph{Nucl.\ Phys.} \textbf{\bf B 534} (1998) 467.
\bibitem{kleefeld2} F.\ Kleefeld,
\emph{Doctoral Thesis} (University of Erlangen-N\"urnberg, 1999).
\bibitem{kleefeld3} F.\ Kleefeld, \emph{Acta Physica Polonica} \textbf{\bf B30} (1999) 981.
\bibitem{kleefeld4} F.\ Kleefeld, ``Consistent Effective Description of Nucleonic Resonances in an Unitary Relativistic Field-Theoretic Way'', in \emph{Relativistic Nuclear Physics and Quantum Chromodynamics} (2 volumes), 
Proc.\ XIV Int.\ Seminar on High Energy Physics Problems (ISHEPP 98),
17.--22.8.1998, JINR, Dubna, Russia, edited by A.M. Baldin, V.V. Burov, 
\copyright 2000 by Joint Institute for Nuclear Research, Dubna
(ISBN 5-85165-570-4 (Vol.\ I)), pp.\ 69-77 in Vol. I (nucl-th/9811032).
\bibitem{wick1} G.C.\ Wick, A.S.\ Wightman, E.P.\ Wigner, \emph{Phys.\ Rev.} \textbf{\bf 88} (1952) 101.
\bibitem{jun1} Jun-Chen Su, hep-th/9805195, hep-th/9805192, hep-th/9805193, hep-th/9805194. 
\bibitem{hoo1} G.\ 't Hooft, \emph{Nucl.\ Phys.} \textbf{\bf B 35} (1971) 167. 
\bibitem{kleefeld5} F.\ Kleefeld, ``Does it make any sense to talk about a $\Delta$-isobar?'', in Proc. 2002 CFIF Fall Workshop on \emph{Nuclear Dynamics: from Quarks to Nuclei}, 31.10.--2.11.2002, Lisbon, Portugal, nucl-th/0212008.  
\bibitem{qcd1} \emph{Handbook of QCD} (3 volumes), edited by M.\ Shifman, 
                      \copyright 2001 by World Scientific Publishing. 
\bibitem{delbourgo1} R. Delbourgo, M.D.\ Scadron, \emph{Mod.\ Phys.\ Lett.} \textbf{\bf A 10} (1995) 251; M.\ Gell-Mann, M.\ L\'{e}vy, \emph{Il Nuovo Cimento} \textbf{\bf 16} (1960) 705; M.\ L\'{e}vy, \emph{Il Nuovo Cimento} \textbf{\bf A 52} (1967) 23. 
\bibitem{alk1} R.\ Alkofer, H.\ Reinhardt, 
                      \emph{Chiral Quark Dynamics} 
                      (Lecture Notes in Physics; New Series m: Monographs; Vol.\ 33 (= m 33)), 
                      \copyright 1995 by Springer-Verlag Berlin Heidelberg. 
\bibitem{haymaker1} R.W.\ Haymaker, J.\ Perez-Mercader, \emph{Phys.\ Rev.} \textbf{\bf D 27} (1983) 1353 and \emph{Phys.\ Lett.} \textbf{\bf 106 B} (1981) 201. 
\bibitem{dass1} N.D.\ Hari Dass, V.\ Soni, \emph{Phys.\ Rev.} \textbf{\bf D 65} (2002) 095005. 
\bibitem{hoo4} G.\ 't Hooft, \emph{Phys.\ Rep.} \textbf{\bf 142} (1986) 357. 
\bibitem{scadron1} M.D.\ Scadron, F.\ Kleefeld, G.\ Rupp, E.\ van Beveren, hep-ph/0211275. 
\bibitem{beveren1} E.\ van Beveren, F.\ Kleefeld, G.\ Rupp, M.D.\ Scadron, \emph{Mod.\ Phys. Lett.} \textbf{\bf A 17} (2002) 1673. 
\bibitem{andrianov1} A.A.\ Andrianov, D.\ Espriu, \emph{J.\ HEP} \textbf{\bf 10} (1999) 022; J.\ Bijnens, \emph{Phys.\ Rep.} \textbf{\bf 265} (1996) 369.
\bibitem{hoo2} G.\ 't Hooft, \emph{Nucl.\ Phys.} \textbf{\bf B 72} (1974) 461, \textbf{\bf B 75} (1974) 461; 
E.\ Witten, \emph{Nucl.\ Phys.} \textbf{\bf B 160} (1979) 57.
\bibitem{bicudo1} P.J.A.\ Bicudo, J.E.F.T.\ Ribeiro, A.V. Nefediev, \emph{Phys.\ Rev.} \textbf{\bf D 65} (2002) 085026.
\bibitem{rembiesa1} P.\ Rembiesa, \emph{Phys.\ Rev.} \textbf{\bf D 38} (1988) 1916. 
\bibitem{okubo1} S.\ Okubo, V.S.\ Mathur, \emph{Phys.\ Rev.} \textbf{\bf D 1} (1970) 2046; P.\ Carruthers, R.W.\ Haymaker, \emph{Phys.\ Rev.} \textbf{\bf D 4} (1971) 1808 and \emph{Phys.\ Rev.} \textbf{\bf D 6} (1972) 1528; J.J.\ Brehm, \emph{Phys.\ Rev.} \textbf{\bf D 9} (1974) 1818. 
\bibitem{bramon1} A.\ Bramon, Riazuddin, M.D.\ Scadron, \emph{J.\ Phys.} \textbf{\bf G 24} (1998) 1. 

\end{thebibliography}
{
}

\end{document}